\documentclass[conference,10pt]{IEEEtran}
\usepackage[nolist]{acronym}
\usepackage{amsmath}
\usepackage{amsthm}
\usepackage{amssymb}
\usepackage{cases}
\usepackage{enumitem}
\usepackage{balance}
\usepackage{graphicx}
\usepackage{siunitx}
\theoremstyle{definition}

\usepackage[ruled,vlined]{algorithm2e}
\SetKwInOut{Input}{Input}
\SetKwInOut{Output}{Output}
\SetAlFnt{\footnotesize} % change algorithm font size
\LinesNumbered  % add the line numbers
\SetAlCapSty{xAlCapSty}

\usepackage{subcaption}
\usepackage[labelformat=parens,labelsep=quad, skip=3pt]{caption}
\usepackage{blindtext}   %%%%TAB
\usepackage{booktabs,tabularx}
\DeclareMathOperator{\E}{\mathbb{E}}
\usepackage[table,xcdraw]{xcolor}
\DeclareMathOperator*{\argmax}{arg\,max}
\newcommand{\figref}[1]{\figurename~\ref{#1}}
\newcommand\norm[1]{\left\lVert#1\right\rVert}

\begin{document}
\bstctlcite{IEEEexample:BSTcontrol}

\title{Deep Reinforcement Learning for Dynamic Band Switch in Cellular-Connected UAV}
\author{\IEEEauthorblockN{Gianluca Fontanesi\IEEEauthorrefmark{1}, Anding Zhu\IEEEauthorrefmark{1} and Hamed Ahmadi\IEEEauthorrefmark{1}\IEEEauthorrefmark{2}}
\\
\IEEEauthorrefmark{1}School of Electrical and Electronic Engineering, University College Dublin, Ireland\\
\IEEEauthorrefmark{2}Department of Electronic Engineering, University of York, Heslington, York YO10 5DD, United Kingdom
}

\maketitle

\begin{abstract} 
The choice of the transmitting frequency to provide cellular-connected Unmanned Aerial Vehicle (UAV) reliable connectivity and mobility support introduce several challenges. Conventional sub-6 GHz networks are optimized for ground Users (UEs).
Operating at the millimeter Wave (mmWave) band would provide high-capacity but highly intermittent links.
To reach the destination while minimizing a weighted function of traveling time and number of radio failures, we propose in this paper a UAV joint trajectory and band switch approach.
By leveraging Double Deep Q-Learning we develop two different approaches to learn a trajectory besides managing the band switch. A first blind approach switches the band along the trajectory anytime the UAV-UE throughput is below a predefined threshold. In addition, we propose a smart approach for simultaneous learning-based path planning of UAV and band switch. The two approaches are compared with an optimal band switch strategy in terms of radio failure and band switches for different thresholds. Results reveal that the smart approach is able in a high threshold regime to reduce the number of radio failures and band switches while reaching the desired destination.
\end{abstract}

\begin{acronym} 
\acro{3G}{Third Generation}
\acro{4G}{Fourth Generation}
\acro{5G}{Fifth Generation}
\acro{6G}{Sixth Generation}
\acro{3GPP}{3rd Generation Partnership Project}
% \acro{ACO}{Ant Colony Optimization}
% \acro{A2G}{Air-to-Ground}
\acro{BB}{Base Band}
\acro{BBU}{Base Band Unit}
\acro{BER}{Bit Error Rate}
\acro{BH}{Backhaul}
\acro{BPP}{Binomial Point Process}
\acro{BS}{Base Station}
\acro{BW}{bandwidth}
\acro{C-RAN}{Cloud Radio Access Networks}
\acro{CAPEX}{Capital Expenditure}
\acro{CDF}{Cumulative Distribution Function}
\acro{CoMP}{Coordinated Multipoint}
\acro{CPRI}{Common Public Radio Interface }
\acro{CU}{Centralized Unit}
% \acro{C$\&$C}{Command and Control}
\acro{D2D}{Device-to-Device}
\acro{DAC}{Digital-to-Analog Converter}
\acro{DAS}{Distributed Antenna Systems}
\acro{DBA}{Dynamic Bandwidth Allocation}
\acro{DDQN}{Double Deep Q-Learning}
\acro{DL}{Downlink}
\acro{DNN}{Deep Neural Network}
\acro{DQN}{Deep Q-Network}
\acro{DDQN}{Double DQN}
\acro{DU}{Distributed Unit}
\acro{FBMC}{Filterbank Multicarrier}
\acro{FEC}{Forward Error Correction}
\acro{FH}{Fronthaul}
\acro{FL}{Federated Learning}
\acro{FFR}{Fractional Frequency Reuse}
\acro{FSO}{Free Space Optics}
% \acro{GA}{Genetic Algorithms}
% \acro{G2A}{Ground-to-Air}
\acro{GS}{Ground Station}
\acro{GSM}{Global System for Mobile Communications}
\acro{HAP}{High Altitude Platform}
\acro{HetNet}{Heterogeneous Network}
\acro{HL}{Higher Layer}
\acro{HARQ}{Hybrid-Automatic Repeat Request}
\acro{KPI}{Key Performance Indicator}
\acro{kg}{Kilogramm}
\acro{IoT}{Internet of Things}
\acro{LAN}{Local Area Network}
\acro{LAP}{Low Altitude Platform}
\acro{LL}{Lower Layer}
\acro{LoS}{Line of Sight}
\acro{LTE}{Long Term Evolution}
\acro{LTE-A}{Long Term Evolution Advanced}
\acro{MAC}{Medium Access Control}
\acro{MAP}{Medium Altitude Platform}
\acro{MC}{Monte Carlo}
\acro{MDP}{Markov Decision Process}
\acro{ML}{Medium Layer}
\acro{MME}{Mobility Management Entity}
\acro{mmWave}{millimeter Wave}
\acro{MIMO}{Multiple Input Multiple Output}
\acro{ML}{Machine Learning}
\acro{MSE}{Mean Square Error}
\acro{NFP}{Network Flying Platform}
\acro{NFPs}{Network Flying Platforms}
\acro{NN}{Neural Network}
\acro{NLoS}{Non-Line of Sight}
\acro{RL}{Reinforcement Learning}
\acro{NR}{New Radio}
\acro{OFDM}{Orthogonal Frequency Division Multiplexing}
\acro{PAM}{Pulse Amplitude Modulation}
\acro{PAPR}{Peak-to-Average Power Ratio}
\acro{PDF}{Probability Density Function}
\acro{PER}{Prioritized Experience Replay}
\acro{PGW}{Packet Gateway}
\acro{PHY}{physical layer}
\acro{PP}{Poisson Process}
\acro{PSO}{Particle Swarm Optimization}
\acro{PTP}{Poin to Point}
\acro{QAM}{Quadrature Amplitude Modulation}
\acro{QoE}{Quality of Experience}
\acro{QoS}{Quality of Service}
\acro{QPSK}{Quadrature Phase Shift Keying}
\acro{RAU}{Remote Access Unit}
\acro{RAN}{Radio Access Network}
\acro{RB}{Resource Block}
\acro{RF}{Radio Frequency}
\acro{RN}{Remote Node}
\acro{RRH}{Remote Radio Head}
\acro{RRU}{Remote Radio Unit}
\acro{RRC}{Radio Resource Control}
\acro{RRU}{Remote Radio Unit}
\acro{RSS}{Received Signal Strength}
\acro{RSRP}{Reference Signals Received Power}
\acro{RU}{Remote Unit}
\acro{SCBS}{Small Cell Base Station}
\acro{SDN}{Software Defined Network}
\acro{SINR}{Signal-to-Noise-plus-Interference Ratio}
\acro{SIR}{Signal-to-Interference Ratio}
\acro{SNR}{Signal-to-Noise Ratio}
\acro{SON}{Self-organising Network}
\acro{TETRA}{Trans-European Trunked Radio}
\acro{TD}{Temporal Difference}
\acro{TDD}{Time Division Duplex}
\acro{TD-LTE}{Time Division LTE}
\acro{TDM}{Time Division Multiplexing}
\acro{TDMA}{Time Division Multiple Access}
\acro{UE}{User Equipment}
\acro{UL}{Uplink}
\acro{UAV}{Unmanned Aerial Vehicle}
\acro{UPA}{Uniform Planar Square Array}

\end{acronym}

\section{Introduction}
Integrating \acp{UAV} into cellular communication systems as \acp{UE} is envisioned as an effective solution to support the \ac{UAV}'s mission specific rate-demanding data communication while improving the robustness of the \ac{UAV} navigation \cite{zeng2018cellular}.
% Ground cellular \acp{BS} are densely deployed and are potentially able to provide high data rate and efficient \ac{UAV} monitoring. 
This vision of cellular connected \acp{UAV} communication, however, poses new research challenges due to the significant differences from conventional communication systems.
\acp{UAV}-\ac{UE} have typically higher altitude, higher mobility and have more stringent constraints on the power and operational time than the corresponding ground ones \cite{mozaffari2019tutorial}.
In addition, the existing cellular network operating at sub-6 GHz is bandwidth limited and perform poorly at high \ac{UAV} heights, due to the interference perceived from the down-tilted antennas at the ground \acp{BS} \cite{geraci2018understanding}.
% In particular, providing a reliable communication link from the ground \ac{BS} during the \ac{UAV}'s trajectory is still object of investigation.
% has a drastic impact on the successful completion of the \ac{UAV} mission.
% has various drawbacks on the performance of the link to the \ac{UAV}. As first, 
As a consequence, during its trajectory, a \ac{UAV} is very likely to experience radio link failures. The \acp{UAV}' mobility and flexibility offer a degree of freedom to circumvent these issues.
The \ac{UAV} path design that aims to respect a quality-of-connectivity constraint and minimize the travelling time goes under the name of \textit{Communication-aware} trajectory.
Several works have optimized the \ac{UAV} trajectory under connectivity constraints using graph based \cite{zhang2019trajectoryOutage} or dynamic programming based solutions \cite{bulut2018trajectory}.
The above traditional optimization solutions are time consuming and computationally complex.
For this reason, \ac{RL} approaches have been recently investigated. Compared to a traditional optimization approach, \ac{RL} is able of making decisions interacting iteratively with the environment. 
% This allows the \ac{UAV} to change and adapt real time its trajectory while maintaining the specific connectivity requirement.
A double Q-Learning approach is proposed in \cite{khamidehi2020double} to solve a joint trajectory and outage time constraint problem. A \ac{TD} learning method is utilized in \cite{zeng2019path} to design the \ac{UAV}-\ac{UE} trajectory while minimizing the mission completion time and the disconnection duration.

Besides, exploiting their mobility, \acp{UAV} can establish short \ac{LoS} communication links, that represents an ideal situation to transmit at \ac{mmWave} band.
A \ac{mmWave} link offers a wide spectrum and enables the use of directional beamforming, providing high data rate \cite{susarla2020learning}. The work in \cite{susarla2020learning} addresses the cellular connected \ac{UAV} communication-aware trajectory problem learning simultaneously the \ac{mmWave} beam and trajectory via \ac{DQN} to improve the \ac{UL} performance. 
However, due to the severe attenuation and sensitivity to blockages, \ac{mmWave} links are highly intermittent, leading to frequent radio failures at low \ac{UAV} altitudes \cite{fontanesi2020outage}.

A promising solution to improve the system robustness is to support two different frequency ranges, leading to integrated dual mode sub-6 GHz/\ac{mmWave} systems \cite{semiari2019integrated}. 
Dual connectivity would allow to exploit the complementary advantages of both the frequency bands and reduce consistently outage status through band switch algorithms.
Some literature has recently studied the band switch problem for conventional ground \acp{UE} in a 2D environment. Works in \cite{burghal2019deep}-\cite{mismar2020deep} propose \ac{ML} classifiers that select the best band based on previous rate measurements within a temporal window.
Within the \ac{UAV} communication context, in \cite{ghazzai2018trajectory} \acp{UAV} are equipped with dual band \ac{mmWave} and sub-6 GHz communication modules to act as relays to minimize the total service time, but the switch algorithm between the bands is not investigated.

In this paper, differently from previous works, we aim to solve the communication aware 3D trajectory problem of \ac{UAV}-\ac{UE} proposing a learning strategy to both manage the trajectory and the band switch policy. We firstly formulate a trajectory problem in order to minimize the travelling time and radio failures of the \ac{UAV}. Then we propose a 3D DQN based algorithm to dynamically solve it adjusting the position and the operating frequency of the \ac{UAV}-\ac{UE}. Finally, we present also a blind switch strategy and introduce a baseline optimal band switch policy. We compare the performance of these three approaches in term of number of radio failures and band switches in the result section. We observe that the smart approach significantly improves the performance of the ground \ac{BS} link for stringent thresholds.

\section{System Model}\label{SystemModel}
%In this section we introduce the main parts of the system model.%, including network topology (Section \ref{NetworkTopology}), channel propagation model in Section \ref{ChannelModel}, and the communication model (Section \ref{CommunicationModel}).
\subsection{Network Topology} \label{NetworkTopology}
Consider the downlink of a dual band wireless cellular network, that operates in two frequency bands with frequency $f_c$ and bandwidth $B_c$, $c \in \{1,2\}$.
A set of ground \acp{DU} is deployed in an area $\mathcal{X} \in \mathbb{R}^3$, connected to a single \ac{CU}, and serve a set of $N$ \ac{UAV}-\acp{UE}.
Both \acp{DU} and \acp{UE} are equipped with interfaces which allow them to transmit at both frequency bands.  
We assume the transmission occurs in a single frequency at a time, generating inter-cell interference to neighboring cells working at the same frequency band.
Each \ac{DU} is assumed to have $n_b$ \acp{RB} and transmits with same power $P_{DU}^c$. 
Without loss of generality, we focus on a single link to a \ac{UAV}-\ac{UE} (hereafter addressed as UAV).

\subsection{Ground-to-Air Channel Model} \label{ChannelModel}
In this section we present the path loss, antenna and fading models, considering $f_1$ a lower sub-6 GHz band and $f_2$ the \ac{mmWave} band.
\subsubsection{Path Loss and Antenna Model}
The ground-to-air channel model is subject to \ac{LoS}/\ac{NLoS} variations based on the building distribution in $\mathcal{X}$ and \ac{UAV} height that lead to 
blockages.
For modeling the path loss $L^1(d)$ at sub-6 GHz we consider the urban Macro (UMa) path loss specified in \ac{3GPP} \cite{3GPPStudy_LTE_aerial}.
The path loss $L^2(d)$ at \ac{mmWave} follows a path loss model \cite{fontanesi2020outage}:
\begin{equation}\label{eq:pathLoss_f2}
 L^2(d)=
    \begin{cases}
       l^{L}(d)= X_L d^{-\alpha_L};\\
       l^{NL}(d)= X_{NL} d^{-\alpha_{NL}};
    \end{cases}
\end{equation}
where $d$ is the ground BS-UAV distance, parameters $\alpha_L$, $\alpha_{NL}$ and $X_L$, $X_{NL}$ represent, respectively, the path loss exponent for \ac{LoS}/\ac{NLoS} and the path loss at 1 meter distance.
% \subsubsection{Antenna Model}
Each \ac{DU} has three sectors separated by $\ang{120}$ as characterized by 3GPP specification \cite{3GPPStudy_LTE_aerial}. Each sector is equipped with a vertical $N_1$-element uniform linear array (ULA) at $f_1$ and a Uniform Planar Array (UPA) $N_2 \times N_2$ at $f_2$ tilted with angle $\phi_1$ and $\phi_2$. We denote the total ground \ac{DU} directional radiation pattern as $G(\Theta, \phi)$, where $\Theta$, $\phi$ are elevation and azimuth angles.
% Each antenna element has directivity of $A_E(\theta, \phi)$, where $\theta$ and $\phi$ are the spherical angles in local coordinate system of the origin at the antenna location.
We assume the \ac{UAV} is equipped with a conventional omni-directional antenna of unitary gain in any direction to maintain low complexity and cost at the \ac{UAV}.

\subsubsection{Fading Model}
We model the small scale fading fading power $f_{0,i}^2$ with $i$ $\{LoS, NLoS\}$ as a Nakagami-m fading model, that covers a wide range of fading environments, including both sub-6 GHz ($m=1$) and \ac{mmWave} ($m=3$). Accordingly, $f_0^2$ follows a Gamma distribution with $\E[f_0^2] = 1$.

\subsection{UAV Mobility Model}
The \ac{UAV} moves at constant speed $V=V_{max}$ along a 3D trajectory of duration $T$ that can be divided into $K$ discrete segments with interval $\delta_k = T/K$, $k = \{1,...,K\}$. 
$\delta_k$ is chosen arbitrarily small so that within each step the large scale signal power received by the UAV remains approximately unchanged. Each segment is thus described by its discrete coordinates $\mathbf{q_k}= [x_k, y_k, h_k]$.
The trajectory of the \ac{UAV} starts from a random position $\mathbf{q_I} \in \mathcal{X}$ in the given area of interest $\mathcal{X}$, delimited by borders $[x_{min},x_{max}]\times[y_{min},y_{max}]$ while the altitude of the \ac{UAV} must satisfy $h_k \in [h_{min},h_{max}]$, where $h_{min}$ represents the lower bound of UAVs’ altitude to avoid collisions and $h_{max}$ represents the upper bound.
The trajectory ends when the \ac{UAV} reaches a final destination $\mathbf{q_F}$. % \pm \Delta_f$. 

\subsection{Communication Model} \label{CommunicationModel}
The achievable rate $R_{A}^c$ at $\mathbf{q_k}$ in band $c$ can be expressed as:
$
    R_{A}^c(\mathbf{q_k})= B_c \E (log_2 (1+SINR^c(\mathbf{q_k}))),
$
where $B_c$ is the bandwidth assigned to the link and $SINR^c$ the \ac{SINR}.  
At each trajectory step $\mathbf{q_k}$, the \ac{UAV} associates with the m-th \ac{DU} at position $\mathbf{b}_m = [x_m , y_m , h_m ]$ providing the highest average \ac{RSRP} at the current frequency.
Assuming that the associated cell remains unchanged within each trajectory step, the \ac{SINR} at the \ac{UAV} can be denoted as:
\begin{equation}
   SINR^c(\mathbf{q_k},m) = \frac{\gamma_m^c(k)}{\sigma^2 + \sum_{j \neq m} \gamma_j^c(k)}, 
\end{equation}
where $\gamma_m^c(k)$ is the received power $\gamma_m^c(k)= P_{DU}L^c(\norm{\mathbf{b}_m-\mathbf{q}_k}_2) f_0^2(\mathbf{q}_k)G(\Theta, \phi)$ and terms in the denominator are, respectively, the spectral noise and the interference.
% where $f$ is the fading channel coefficient, and $G$ the DU antenna gain. .
Note that a subscript to emphasize the $\{LoS, NLoS\}$ dependence of $L()$ and $f$ has been omitted to lighten the notation.

To maintain a reliable connection, \ac{DU}-\ac{UAV} link must satisfy a rate threshold $R_{TH}$.
We define a \textit{radio failure} on the link if $R_{A}^c(\mathbf{q_k}) < R_{TH}$ occurs at step $k$.
In addition, we introduce a failure indicator as:
\begin{equation}\label{eq:radiofailure}
F (\mathbf{q_k}) =
    \begin{cases}
    1, & \mbox{if } R_{A}^c(\mathbf{q_k}) < R_{TH}\\
    0, & \mbox{otherwise}.
\end{cases}
\end{equation}
To avoid a radio failure, at each discrete step $k$ the \ac{UAV} may require the \ac{CU} a band switch to change the transmission frequency. 
% The band switch design is challenging. Due to overhead signaling and procedure, frequent and unnecessary band switches degrade the performance of the system by making frequent downlink rate interruptions.
Prior to formulating the problem, in what follows we introduce the band switch procedure and signalling.

\subsection{Band switch procedure}\label{BandSwitchPolicy}
% Consider the scenario described above depicted in \figref{fig:ScenarioBandSwitch}.
The band switch policy industry standard for a conventional ground \ac{UE} is composed of several iterative steps which can be summarized as follows :  1) if the serving rate $R_{A}^c$ at current frequency drops below threshold $R_{TH}$, the \ac{UE} initiates a band switch procedure with the serving cell, 2)  the \ac{UE} utilizes a time $\Delta_t$ to measure if a cell at target frequency offers a higher \ac{RSRP} than the serving cell, 4) if yes, the \ac{UE} triggers a band switch procedure and reports the measurement back to the \ac{DU}, 5) the \ac{CU} makes a decision on the suitable \ac{DU} at target frequency for band switch, based on the \ac{UE} measurements. %, \cite{yin2019general},
%It should be noted that a standardized band switch procedure in the context of \ac{UAV} communication has not been developed yet. 
Applying the above policy to \ac{UAV} communications would introduce several challenges.
%First, as we will show in Section \ref{SimulationResults}, 
The mobility of the \ac{UAV} might cause deep \ac{RSRP} fluctuations between two consequent trajectory steps $\mathbf{q_k}$ and $\mathbf{q_{k+1}}$. 
This might lead to subsequent band switch procedures and potential ping-pong effect. 
%Frequent switch procedures require additional signaling exchanges and increase the latency and the energy consumption on the \ac{UAV}.
Note that, due to the measurement gap, in case of several band switches the effective throughput at the \ac{UE} suffers a significant reduction \cite{mismar2020deep}.
Thus, we modify the above band switch procedure to adapt it for a \ac{UAV} moving between cells belonging to different \acp{DU} controlled by a central \ac{CU}. 
Here, motivated by the above mentioned challenges, we formulate a real-time trajectory and band switch procedure exploiting the capabilities of a \ac{RL} based policy.
In the next sections we formulate the problem and propose two \ac{DQN} algorithms for joint UAV-UE trajectory and band switch.
 %Note that each band switch come at the expense of a reduction $\beta$ in the effective rate due to the measurement gap.

\subsection{Problem formulation}
The goal of the \ac{UAV} is to reach the destination in the minimum amount of steps while minimizing the number of radio failures.
At the same time, to avoid a radio failure and meet the quality of service requirement, at the beginning of each discrete step $k$, the operating frequency can be switched between $f_1$ and $f_2$.
% Note that reaching the destination in the minimum amount of steps reduce the total mission time, which is desirable. However, minimizing the radio failures increase the number of switches.
% additional steps might be required to avoid weak coverage areas. 
Denoting as $K$ the trajectory steps, $N_{s}$ the number of band switches, % $N_{out}$ the number of detachments. % . 
the above problem can be mathematically formulated as:

\begin{subequations}\label{eq:Optimization}
\begin{align}
    \min_{N_s, K,\mathbf{q}_{k}} \quad & \kappa_1 \sum_{k=1}^K F (\mathbf{q_k}) + \kappa_2 K + \kappa_3 N_s\\ \label{eq:cond1_Opt_problem3}
    \textrm{s.t.} \quad & \mathbf{q}_0 = q_I,\\ \label{eq:cond2_Opt_problem3}
  & \mathbf{q}_{K} = q_F,  %\\ \label{eq:cond3_Opt_problem3}
  %& \norm{q_k -q_{k-1}} = \delta_k V_{max}, \\ \label{eq:cond4_Opt_problem3}
%   & R_{A}^c(k) < R_{TH} \text{ for } k \in K,\\ \label{eq:cond5_Opt_problem3}
  %& h_{min} < h_k < h_{max},\\ \label{eq:}
  %& f \in  \{f_1, f_2\}.
%  &  F_k \quad \text{given by } \eqref{eq:radiofailure}, \text{ for } k \in K.
\end{align}
\end{subequations}
where $\kappa_1,\kappa_2, \kappa_3$ weight the summation of the arguments, \eqref{eq:cond1_Opt_problem3}, \eqref{eq:cond2_Opt_problem3} are the starting and final positions. %, \eqref{eq:cond3_Opt_problem3} denotes the \ac{UAV} flying speed, \eqref{eq:cond4_Opt_problem3} is the altitude constraint of the \ac{UAV}. Constraint \eqref{eq:cond5_Opt_problem3} defines that the transmission cannot occur simultaneously at $f_1$ and $f_2$.
Problem \eqref{eq:Optimization} cannot be readily handled by conventional optimization methods due to the non trivial design of $R_{A}^c(\mathbf{q_k})$, function for each position of the fading, and the antenna model described in Section \ref{SystemModel}.

To tackle this challenging problem, we propose a novel approach by leveraging the \ac{DQN} technique. \ac{DQN} based frameworks are able to solve the joint trajectory and band switching problem by exploring different actions, and without the need of prior knowledge or predefined dataset to train the network.

\section{Band Switching as a Deep Reinforcement Learning Problem}
% We now formulate the problem following a \ac{DQN} framework \cite{mnih2015human}.

% This section gives a brief introduction on the \ac{DQN} framework, used in the proposed band switch algorithm.
In the standard \ac{RL} setting an agent acts in an environment over discrete time steps. 
Given that problem \eqref{eq:Optimization} has been formulated in a discrete form, we can directly map it into a \ac{MDP} 5-tuple $\langle \mathcal{S}, \mathcal{A}, \mathcal{P}, \mathcal{R}, \gamma  \rangle$, with space state $\mathcal{S}$, action space $\mathcal{A}$, state transition probability $\mathcal{P}(s'\vert s,a)$, reward function $\mathcal{R}(s,a,s')$, and discount factor $\gamma \in [0,1)$. 
At each time step $k$, the agent receives an observation \textbf{$o_k$} and selects an action $a_k$, receives reward $r_{k+1}$ from the environment and goes from state $s_k$ to a new state $s_{k+1}$ where the cycle restart. We assume an episodic setup, where an episode starts with time step $0$ and concludes with time step $K$. 
Here the states consist in the 3D location of the \ac{UAV} $\mathbf{q_k}$ and band switch indicator $w_k \in \{0, 1\}$, such as $s_k = \{\mathbf{q_k}, w_k\}$ $\in \mathcal{S}$.
Both the location and the band switch indicator are the input of the neural network.
% (see \figref{fig:networkMod2}).
The agent, placed at the \ac{CU} to overcome the limited onboard computing capacity, selects an action in a set of $N$ actions $\{a_1,...a_N\}$ in the action space $\mathcal{A}$. 
The action can be represented by parameters $a_k = \{\eta_k, w_k\}$ where $\eta_k$ corresponds to the movements of the \ac{UAV} in altitude (up or down) or in the horizontal direction (left-right-forward-back). %We don't consider the case where the \ac{UAV} hovers. 
In addition, if $w_k$ is positive, the \ac{CU} triggers a band switch in the network operating frequency without a measurement gap.
Based on the selected action the \ac{UAV} moves in the desired direction for a distance $\delta_k$ at speed $V_{max}$ reaching the position $\mathbf{q_{k+1}}$. %Finally, the \textbf{reward} $r_k$ can be defined as
% The objective is to predict the immediate \ac{UAV}-direction along with its best transmitting frequency using its current rate measurement.
Our goal is to predict the \ac{UAV} direction along with the transmitting frequency to avoid radio failures and reach the destination using the current rate measurement.
% optimize the \ac{UAV} trajectory to minimize the radio failures and traveling time.
With this aim we define as reward function as:
$
    r_{k} = \lambda_1\mathcal{F}_1+\lambda_2 \mathcal{F}_2 + \lambda_3\mathcal{F}_3.
$
$\mathcal{F}_1$ is a negative cost and forces the \ac{UAV} to reach the destination in the minimum amount of steps. $\mathcal{F}_2$ denotes the cost for entering in low normalized rate regions ($R_{TH}/R_{A}^c(\mathbf{q_k})$), that lead to radio failures. $\mathcal{F}_3$ accounts for the band switches occurred until time step $k$. Note that $\mathcal{F}_3$ is a cost since a high number of band switches is not desirable. %due to the disadvantages mentioned in Section \ref{BandSwitchPolicy}.
Thus, $\lambda_1$, $\lambda_2$, $\lambda_3$ balance the impact of the three factors on the reward function and need to be properly adjusted.

In order to achieve the desired goal, we need to find a policy that maximizes the long-term average of both connectivity and path guidance.
We recall that a policy $\pi(s_k,a_k)$ is defined as the probability of taking action $a_k = a$ in state $s_k = s$, %i.e, $\pi(s_k,a_k) = Pr(a_k =a \vert s_k = s)$, 
through interacting with the environment.
Based on this, the problem is equivalent to find a policy $\pi(s_k,a_k)$ that maximizes a cumulative discounted reward over the long run:
\begin{equation}\label{eq:discountedReward}
    G_k = \sum_{i=0}^{K-1} \gamma^i r_{k+i+1}.
\end{equation}
% while minimizing the number of radio failures and switches. 
The discount factor $\gamma$ in \eqref{eq:discountedReward} regulates the relative importance of future and immediate rewards.

After being formulated as an \ac{MDP}, Problem \eqref{eq:Optimization} can now be solved via different \ac{RL} algorithms. %In the following we show our proposed algorithms based on state of the art \ac{DDQN}.

\subsection{DDQN for UAV-UE Trajectory and Smart Band Switch}\label{SmartSwitch}
In the convention \ac{RL} approach, for a given policy $\pi$, a state-action function $Q(s,a)$ is introduced to represent the expected return after taking action $a_k$ in state $s_k$.
In order to deal with large state/action space dimensions \ac{DQN} was introduced in \cite{mnih2015human} to approximate the optimal Q-function using a \ac{DNN} such that $Q(s,a; \theta) \approx Q(s,a)$ where $\theta$ are the network' weights. 
The transitions $(s,a,r,s')$, with $s'=s_{k+1}$ are stored into a memory of finite capacity size $H$ and randomly picked when performing a Q-value update.
At each iteration, \ac{DQN} is trained to minimize the loss:
 \begin{equation}\label{eq:loss}
    L(\theta) = E[(y_k-Q(s_k,a_k \vert \theta_k)^2)]
\end{equation}
where $y_k$ is the target function given by $y_k = r_k + \gamma \max_a Q(s_{k+1},a \vert \theta^{-})$,
where $\theta^{-}$ is introduced following the \textit{target network mechanism}.
We denote the primary \ac{DNN} network weight matrix and target \ac{DNN} weight matrix as $\theta$ and $\theta^{-}$.
We consider a fully connected \ac{DNN} for both networks and the \ac{DNN} parameters $\theta^{-}$ are updated with the parameters $\theta$ every fixed number of steps $C$ to improve the convergence of the algorithm.
Among the improvements and extensions of the baseline \ac{DQN} algorithm, we implement \ac{DDQN} which reduces the overstimation bias of \ac{DQN} with a simple modification of the update rule. In particular, the target function becomes:
 \begin{equation}\label{eq:targetFunction}
    y_k = r_k + \gamma \max_a Q(s_{k+1}, \argmax Q(s_{k+1},a_{k+1},\theta_k), \theta_k^{-}).
\end{equation}

In Algorithm \eqref{alg:Smartrajectory} we introduce the complete pseudocode.

\begin{algorithm}[t]
\caption{DDQN Algorithm for Band Switch Trajectory}\label{alg:Smartrajectory}
\SetAlgoLined

% \KwInput{}
% \KwOutput{} 
\textbf{Initialize:} maximum number of episodes, the replay memory $H$ with capacity $N$, mini batch size $B$ \;
 the DQN $Q$ network with coefficients $\theta$, the target network $\tilde{Q}$ with coefficients $\theta^{-} = \theta$\;
 Algorithm hyperparameters ($\epsilon$, $\lambda_1$,$\lambda_2$,$\lambda_3$,$\gamma$), Rate Threshold $R_{TH}$\;
% \KwResult{Outage Mao for area X}
% Download the Outage BitMap $O$ from the cellular network and denotes the existing database containing all the measurements\;
 \For{episode = 1,...,Max episode}{
 Initialize $q_0 = \{ \mathbf{q_I}\} \in \mathcal{S}$, $w=0$, set step $k \xleftarrow[]{} 0$\;
  \For{each step of episode}{
  \If{$rand( )  < \epsilon$}
        {select action $a_k$ randomly \;
  \Else{choose action $a_k  = \argmax Q(\mathbf{s_k},a_k,\theta)$}
    }
    Agent execute action $a_k$, observe $\{\mathbf{q_{k+1}},w_k\}$\;
    % \If{$\norm{\mathbf{b}_m-\mathbf{q}_F} < \Delta_f$} 
    \If{$w_k == 1$}{$f_c = !f_c$ (\textbf{Band Switch})}
    \If{Destination reached}
        {Terminate Episode \;
%   \Else{\If{$w_k == 1$}{$f_c = !f_c$ (\textbf{Band Switch})}}
        }
     Store transition $\{ \mathbf{s}_{k}$, $a_k$, $r_k$, $\mathbf{s}_{k+1}$\} in $H$\;
    %  with maximal priority $p_t = \max_{} p_i$
%     \If{$H$ is full}
%         { \For{each step of episode}{Sample transition $j$ based on \;
%         Compute importance-sampling weight;\;
%         Compute TD-error\;
%         Update Transition priority}
%         Update weights of $\theta$ of Q() by minimizing the loss function;\;
%         Set $\theta_{target} = \theta$ every fixed number of steps.
%   \Else{choose action $a = \argmax Q(q_n,v,\theta)$ (exploitation)}
%       }
    Compute reward $r_k$\;
    Randomly sample mini batch from replay memory $H$\;
    % Perform a gradient descent $Z^{\kappa}(r_{k} + \alpha \max Q^*(s_{k},a'))$\;
    Perform a gradient descent using \eqref{eq:loss} and \eqref{eq:targetFunction}\;
    Update $\tilde{Q} = Q$ every $C$ steps\;
    }
 }
\end{algorithm}

\subsubsection{Analysis of the Smart Switch Algorithm}
The proposed algorithm aims to solve the challenges of the state of the art band switch policy presented in Section \ref{BandSwitchPolicy}. 
After the initialization procedures, the agent performs an initial exploration of the state space through an $\epsilon$-greedy policy (lines ($7-12$) in Algorithm \eqref{alg:Smartrajectory}).
The UAV moves in a custom environment consisting of the area under consideration $\mathcal{X}$ covered by the dual band network, the possible UAV locations and discrete action space. 
At each episode, the environment resets the source \ac{UAV} location, computes new observations and rewards based on the desired action until it reaches destination. 
Since the band switch indicator is included as input state of the \ac{DNN}, goal of the agent is to learn by experience in which states bandswitching is beneficial, without the need for measurement of gaps.

Next, we introduce a Blind DQN Algorithm for the joint UAV-UE Trajectory and Band Switch and an Optimal Algorithm.

\subsection{DDQN for joint UAV-UE Trajectory and Blind Band Switch}
To solve the optimization problem \eqref{eq:Optimization} we propose also a Blind algorithm, that focuses on learning the best trajectory to reach the destination and switches the transmitting frequency any time a radio failure occurs. Differently from the Algorithm presented in Section \ref{SmartSwitch}, the states consist of the 3D location of the \ac{UAV} and the action in the direction only. 
In addition, in the reward function, $F_3 = 0$.
The Algorithm pseudocode is omitted here since it is similar to the one presented in Algorithm \ref{alg:Smartrajectory}.
This band switch algorithm without information about the instantaneous rate at both frequencies, cannot guarantee that the throughput after the band switch will be higher than the original one.

\subsection{Optimal DDQN for UAV-UE Trajectory}
Finally, as benchmark, we introduce an Optimal Algorithm. At each trajectory step $\mathbf{q_k}$ the agent knows the instantaneous achievable rate of the two different bands.
Therefore, the \ac{CU} at each step coordinates the \acp{DU} to transmit at the best frequency, minimizing the radio failures due to a wrong choice about the band.

\section{Simulation Results} \label{SimulationResults}
In this section we evaluate the performance of the proposed \ac{DQN} algorithms and we benchmark them with the Optimal Policy.
\begin{table}[t]
  \centering
  \scalebox{0.9}{
\begin{tabular}{||p{1.5cm}|p{3.5cm}|p{2cm}||}
 \hline
 Parameter & Description & Value\\
 \hline
%  $V_{max}$ & UAV speed  & 20 m/s \\
  $h_{max}$/$h_{min}$ & UAV Max/Min Height & 120/60 m \\
  $N_1$/$N_2 \times N_2$ & Antenna Element $f_1$/$f_2$ & 8/64\\
  $\phi_1$/ $\phi_2$ & Antenna Tilt $f_1$/$f_2$ & -10/10  $\deg$ \\
  $\sigma^2$ &  Noise Power $f_1$/$f_2$ &  -204/-120 db/Hz \\
  $\delta_k$ & time step length & 0.5 s\\
  $\epsilon$  & $\epsilon$-greedy variable & 0.4\\
  $K$ & UAV Max Moves & 200\\
  $\gamma$ &  discount factor & 1 \\
  $H$ & Replay Memory Size & 100000\\
  $\lambda_1$, $\lambda_2$, $\lambda_3$ & Rewards weights & 0.1, 0.8, 0.1\\
%  $B_2$  & Bandwidth mmWave & 1800KHz & &  &\\
 \hline
\end{tabular}}
\caption{Simulation parameters}
\label{table:SimulationParameter}
\end{table}
We consider a deployment of 5 dual band \acp{DU} in an area of side $2$ km where buildings are modeled as for the ITU-R urban statistical model \cite{ITUBlockageModel} with maximum height of $50$m. 
The two frequency bands considered are $f_1 = 2$ GHz and $B_1 = 180$ kHz, and $f_2 = 28$ GHz and $B_2$ = 1800 kHz. We use one \ac{RB} at sub-6 GHz while ten at \ac{mmWave} \cite{mismar2020deep}.
We choose to transmit at $0.1$ W at \ac{mmWave} and at $1$ W at sub-6 GHz. 
In the proposed \ac{DQN} algorithms, the \ac{DNN} consists of input layer, four hidden layers, one output layer, all fully connected feedforward, activated using Rectified Linear Units (ReLU) and trained with Adam optimizer. 
Each training run was conducted using a batch size of $32$ for a minimum of $3500$ iterations. 
% The number of training iterations was determined to be the shortest number of iterations where the training loss for all environments converges approximately to zero. 
Table \ref{table:SimulationParameter} presents all simulation parameters.

\subsection{Simulation Environment}
\begin{figure}[tb]
\centering
\begin{subfigure}{4.6cm}
\centering\includegraphics[width=4.6cm]{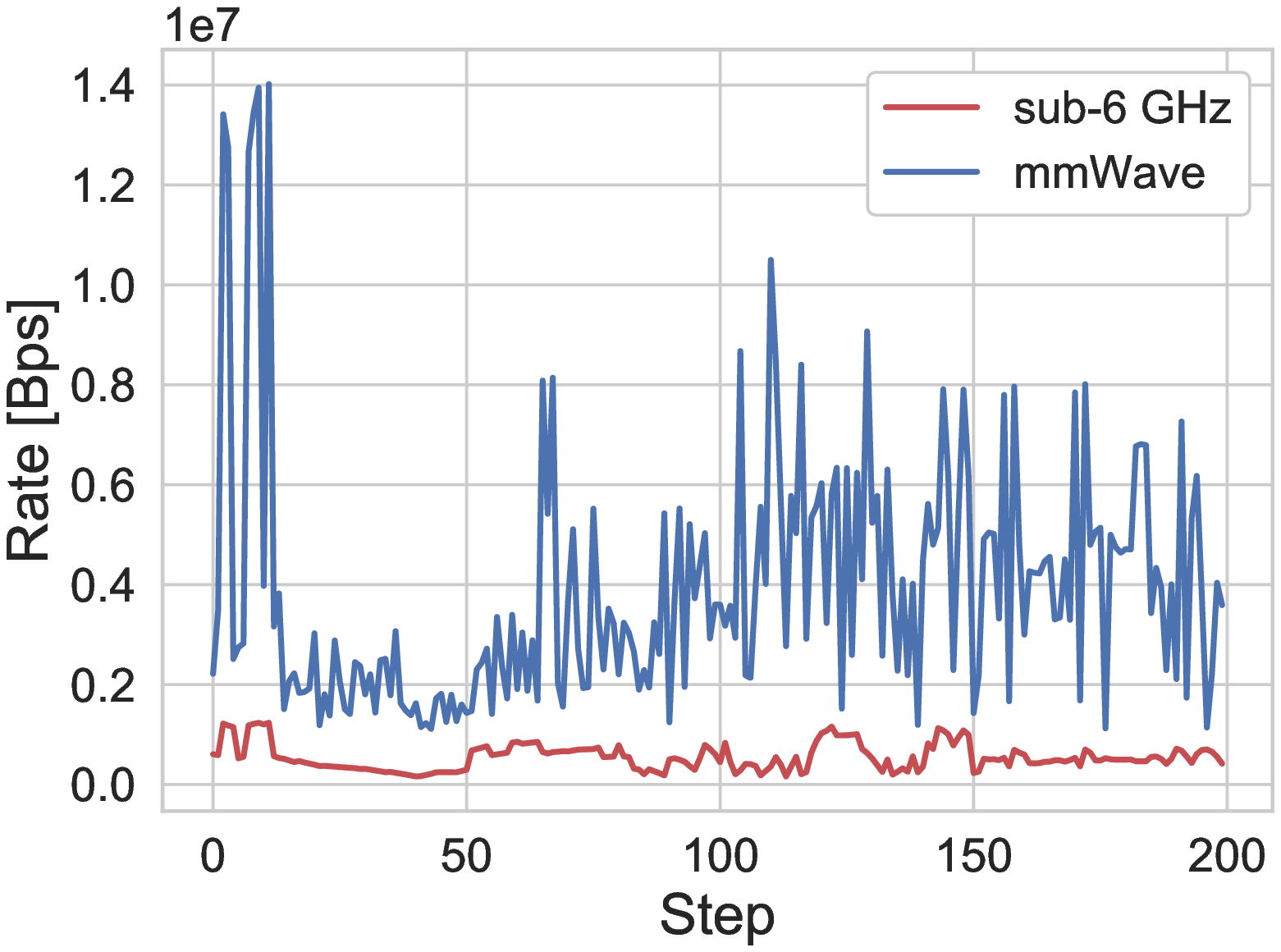}
\caption{}\label{fig:RateBehaviour}
\end{subfigure}%
\begin{subfigure}{4.15cm}
\centering\includegraphics[width=4.15cm]{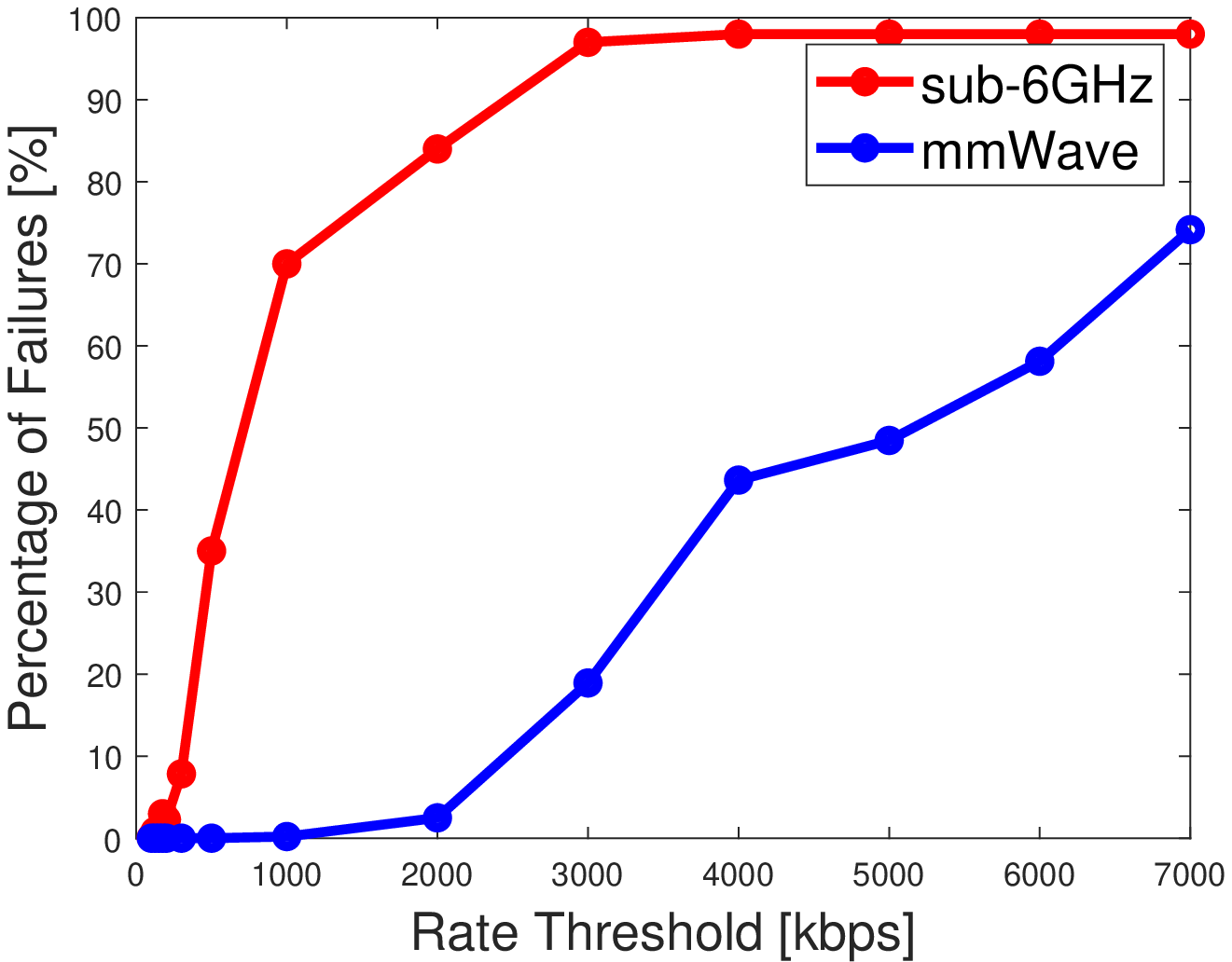}
\caption{}\label{fig:failures_3D}
\end{subfigure}
\caption{Evolution of the \ac{UAV}-\ac{UE} rate at sub-6 GHz and mmWave along one single trajectory episode (a). Samples are collected at each step k, according to the framework described in Section \ref{CommunicationModel}. In (b) we show the Radio Failures for a 3D trajectory at sub-6 GHz and mmWave}
\label{fig:}
\end{figure}

In \figref{fig:RateBehaviour}, we show the rate trend along with the same trajectory episode at sub-6 GHz and \ac{mmWave}.
The rate has sudden drops both at sub6GHz and \ac{mmWave}, due to the presence of fading and obstacles. In particular, at \ac{mmWave}, the rate can be very high or very low, due to the higher bandwidth and, at the same time, high sensitivity to blockages.
In \figref{fig:failures_3D} we show the total percentage of detachments completing the \ac{UAV} missions at one frequency only, sub-6 GHz or \ac{mmWave} band, and without any band switch mechanism. 
Intuitively, the number of failures increases as we increase the rate threshold.
Both the above results motivate the band switching approach presented in this paper.

\subsection{Band Switch Policies}
\begin{figure}
     \centering
     \begin{subfigure}[b]{4.25cm}
         \centering
         \includegraphics[width=4.25cm]{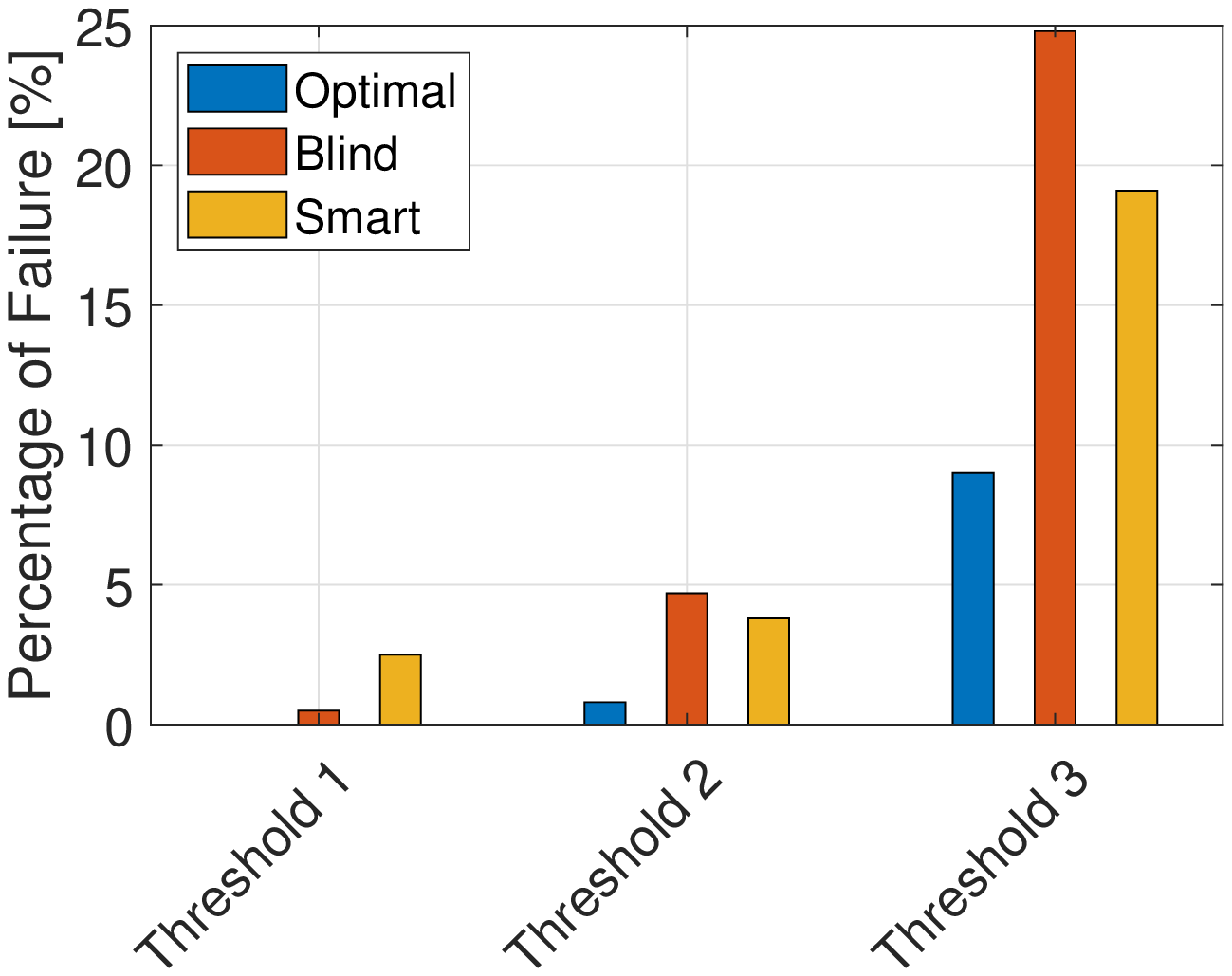}
    \caption{Radio Failures for the Blind and Smart approaches, compared with an Optimal policy}
    \label{fig:bar_RadioFailures}
     \end{subfigure}
     \hfill
     \begin{subfigure}[b]{4.45cm}
         \centering
         \includegraphics[width=4.45cm]{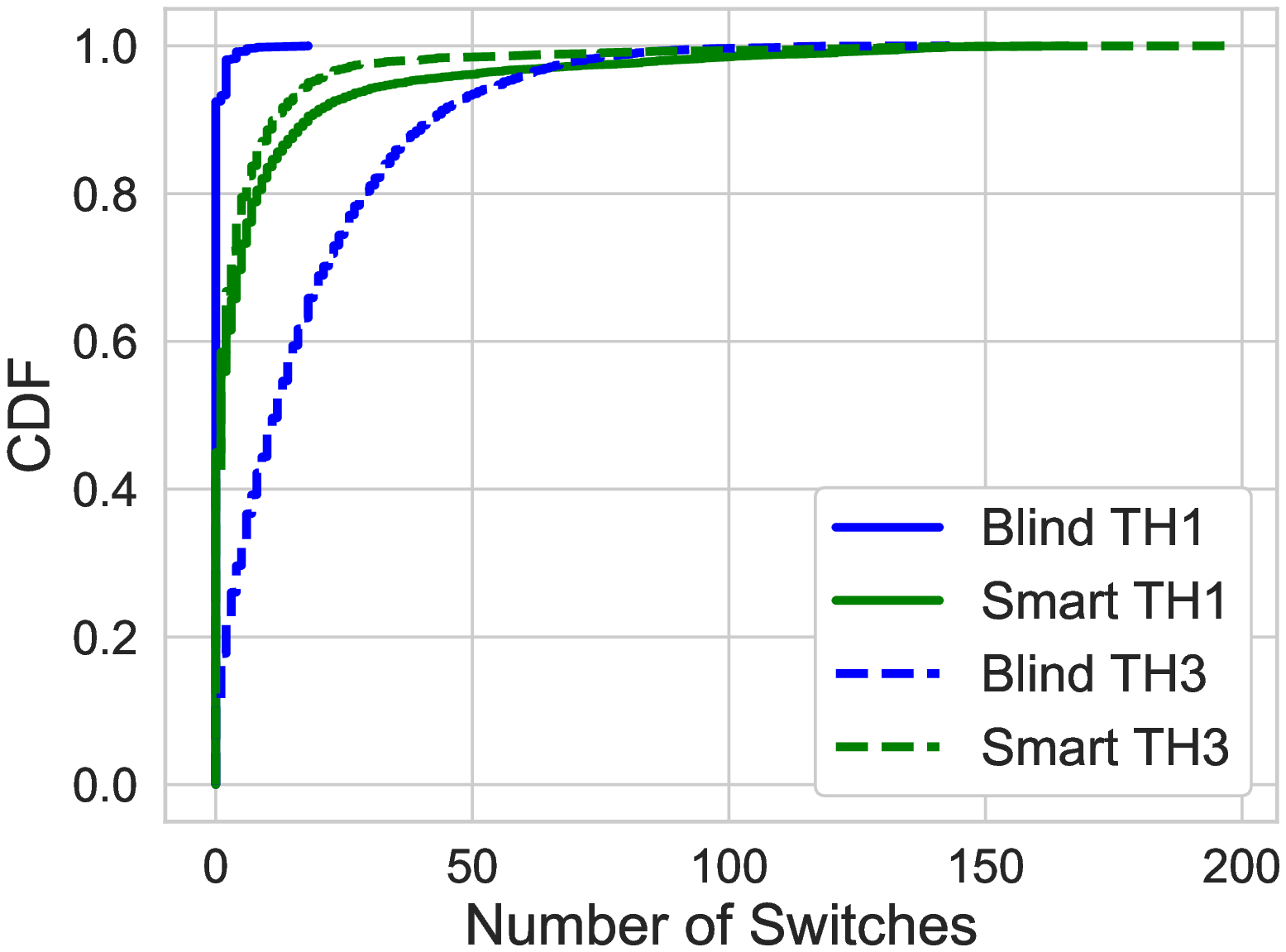}
    \caption{CDF of number of switches for the Blind and Smart Algorithm with thresholds 1 and 3}
    \label{fig:cdf_switches}
     \end{subfigure}
        \caption{Performance of the band switch algorithms}
        \label{fig:PerformanceBandSwitch}
\end{figure}

The \ac{UAV}-\ac{UE} starts its trajectory at sub-6 GHz and can switch to the 28 GHz band. We consider a low threshold of $150 \text{Kbps}$ at sub-6 GHz and $3 \text{Mbps}$ at \ac{mmWave} (Treshold 1), a threshold of $300 \text{Kbps}$ at sub-6 GHz and $3 \text{Mbps}$ at \ac{mmWave} (Treshold 2) and a high Threshold 3 of $400 \text{Kbps}$ at sub-6 GHz and $4 \text{Mbps}$ at \ac{mmWave}.
In \figref{fig:bar_RadioFailures} we compare the blind, the smart and optimal switch algorithms for increasing thresholds in term of average percentage of radio failures. 
We observe that at low thresholds, the smart approach performs similarly to the blind approach. 
By increasing the threshold, and as consequence, the probability of a radio failure, we observe a higher increase in the number of radio failures for the blind approach.
While the blind approach switches the band anytime a radio failure occurs, the smart one offers the agent a generic framework that, interacting with the environment, learns to minimize the number of radio failures switching the band only when the target frequency provides higher throughput.
As confirmation of this, in \figref{fig:cdf_switches} we plot the \ac{CDF} of the number of band switches of the blind and smart approach for Threshold 1 and 3. We observe that in a high throughput regime, the smart approach outperforms the blind approach and can significantly reduce the number of switches.

\section{Conclusion}
In this paper, we propose a DQN based approach to solve a communication aware trajectory problem for cellular \ac{UAV}.
In order to minimize the number of radio failures while reaching its target destination, we consider the \ac{UAV} to exploit band switch policies without prior knowledge of the environment.
We define and compare two joint trajectory and band switch \ac{DQN} algorithms, that differentiate based on how the agent decide to switch the transmitting frequency.
We show that a smart switch approach, where the band switch indicator is included as input state of the \ac{DNN}, is able to outperform a blind approach in a high threshold regime.

\section*{Acknowledgment}
{\footnotesize This work was supported by the Irish Research Council under Grant GOIPG$/2017/1741$ and in part by the Science Foundation Ireland under Grant Numbers $13/$RC$/2077$ and $17$/NSFC/$4850$.}
\bibliographystyle{IEEEtran}
{\footnotesize 
\bibliography{References.bib}
}
\end{document}